\input{aipcheck}

\documentclass[
    ,final            
  ]
  {aipproc}

\layoutstyle{6x9}

\def\aj{AJ}%
\def\araa{ARA\&A}%
\def\apj{ApJ}%
\def\apss{Ap\&SS}%
\def\aap{A\&A}%
\def\aaps{A\&AS}%
\def\nat{Nature}%

\begin{document}

\title{The potential for intensity interferometry with $\gamma$-ray telescope arrays}


\author{W.\,J. de Wit}{
  address={School of Physics and Astronomy, University of Leeds, LS2 9JT, UK}
}

\author{S. Le Bohec}{
  address={Department of Physics, University of Utah, 115 S 1400 E, Salt Lake
  City, UT 84112-0830, USA}
}

\author{J.\,A. Hinton}{
  address={School of Physics and Astronomy, University of Leeds, LS2 9JT, UK}
}
\author{R.\,J. White}{
  address={School of Physics and Astronomy, University of Leeds, LS2 9JT, UK}
}
\author{M.\,K. Daniel}{
  address={School of Physics and Astronomy, University of Leeds, LS2 9JT, UK}
}
\author{J. Holder}{
  address={Department of Physics and Astronomy, University of Delaware, Newark,
  Delaware, USA}
}

\begin{abstract}
Intensity interferometry exploits a quantum optical effect in order to
measure objects with extremely small angular scales. The first
experiment to use this technique was the Narrabri intensity
interferometer, which was successfully used in the 1970s to measure 32
stellar diameters at optical wavelengths; some as small as 0.4
milli-arcseconds. The advantage of this technique, in comparison with
Michelson interferometers, is that it requires only relatively crude,
but large, light collectors equipped with fast (nanosecond) photon
detectors. Ground-based $\gamma$-ray telescope arrays have similar
specifications, and a number of these observatories are now operating
worldwide, with more extensive installations planned for the
future. These future instruments (CTA, AGIS, completion 2015) with 30-90 telescopes will
provide 400-4000 different baselines that range in length between 50\,m and a
kilometre. Intensity interferometry with such arrays of telescopes  
attains $50\,\mu$-arcsecond resolution for a limiting $\rm m_{v}\sim8.5$.
Phase information can be extracted from the interferometric measurement with
phase closure, allowing image reconstruction. This technique opens the
possibility of a wide range of studies amongst others, probing the stellar
surface activity and the dynamic AU scale circumstellar environment of stars in
various crucial evolutionary stages. 
Here we focuse on the astrophysical potential of an intensity 
interferometer utilising planned new $\gamma$-ray instrumentation.

\end{abstract}

\maketitle


\section{Introduction}
Optical intensity interferometry (I.I.) offers some considerable advantages over
Michelson interferometry in terms of complexity, cost and wavelength coverage.
Therefore, whilst so far successfully realised for only
one astrophysical instrument (the Narrabri interferometry of Hanbury Brown et al.
\cite{1956Natur.178.1046H}), it has enormous potential for future instruments. Several years ago it was
recognised that the optical requirements for an intensity interferometer match
rather closely those for the arrays of Cherenkov telescopes used in ground-based
$\gamma$-ray astronomy \cite{2006ApJ...649..399L}, namely very large mirrors and modest
optical quality. Furthermore, it seems relatively straightforward
to 'share' the optical systems of such a telescope array between these two
activities. As large arrays of such $\gamma$-ray telescopes are now being planned,
it seems timely to revisit this possibility. The technical details of this
approach are described in a companion paper in these proceedings (Le Bohec et al).
Here we focus on the planned new instrumentation and on the astrophysical
potential of an intensity interferometer utilising such a telescope array.

\section{Cherenkov Telescope Arrays}
The current operational Imaging Atmospheric Cherenkov Telescope (IACT) arrays such as
VERITAS \cite{2006APh....25..391H} and H.E.S.S. \cite{2004NewAR..48..331H}
consist of four 12~m ($\sim100~{\rm m}^{2}$) telescopes compound
mirrors producing on-axis point-spread-functions of $\rm FWHM\approx
0.05^{\circ}$. These telescopes are designed to image the Cherenkov light
produced in the electromagnetic cascades initiated by very-high-energy $\gamma$-rays
(typically $10^{11} - 10^{13}$ electronVolt), and have been used
in the past to do optical measurements.  For example the telescopes of
H.E.S.S. have been used to measure the optical light-curve of the Crab pulsar
\cite{2006APh....26...22H} and to search for ultra-fast optical transients from
binary systems (Deil et al. these proceedings).

Two initiatives currently exist aiming for an order of magnitude improvement in the
sensitivity of IACT arrays. To achieve this, both projects aim to construct a
large array with more and/or larger telescopes, covering an area of approximately 1\,$\rm
km^2$. Larger light collecting dishes, important for I.I. sensitivity, are technically feasible and are currently used by
the single dish MAGIC telescope (17\,m) \cite{2005Ap&SS.297..245C}. A 30\,m
dish is under construction as part of the second phase of the H.E.S.S. project. AGIS
(Advanced $\gamma$-ray Imaging System) \cite{AGIS}
is the American-proposed next generation
$\rm 1km^2$ instrument. The Cherenkov Telescope Array (CTA) \cite{CTA}
is the European equivalent.
The science goals of these projects are extremely diverse, covering
many aspects of both galactic and extragalactic high energy astrophysics as well
as topics from particle physics and cosmology. The left panel of Fig.\,1
illustrates a potential layout of CTA with simulated performance that meets
the goal sensitivity for $\gamma$-ray astronomy \cite{bernloehr}. Some 85 identical
telescopes provide 3570 potential pairings, providing 47 closely-spaced, unique
baselines ranging between 50 and 1000\,m. The idea of using CTA as an intensity 
interferometer is currently under discussion within the consortium.



\begin{figure}
  \includegraphics[height=.275\textheight,width=.275\textheight]{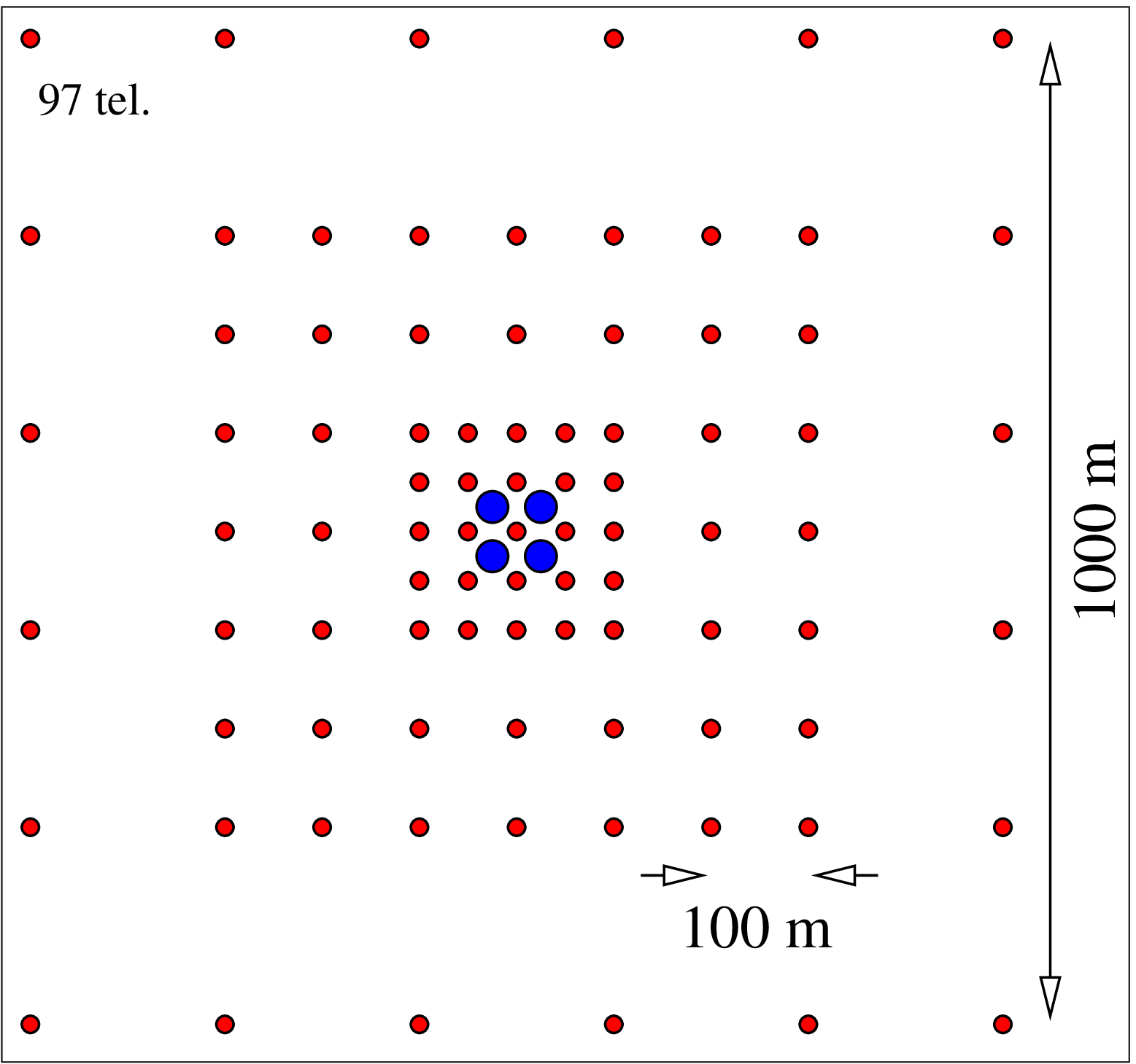}
  \includegraphics[height=.3\textheight,width=.4\textheight]{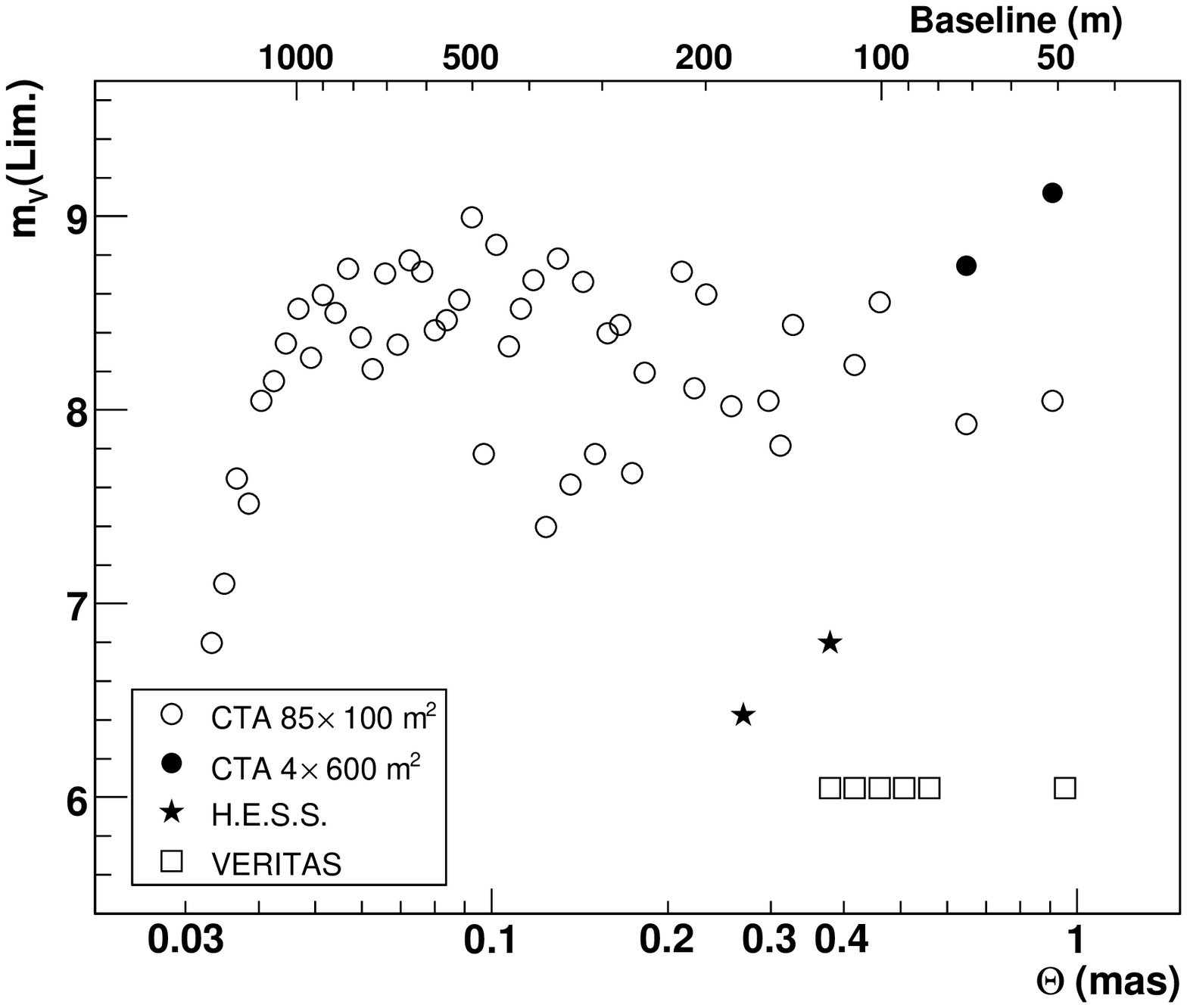}
  \caption{{\it Left:} Proposed lay-out for the future CTA. The small dots are
  the 85 $\rm 100\,m^2$ dishes, the large dots are the 4 $\rm 600\,m^2$
  dishes. {\it Right:} I.I. with
  CTA sensitivity estimate as function of the 47 non-redundant baselines for a 5$\sigma$ detection, in a 5 hours integration on a
  star with 50\% visibility. Final values depend on signal bandwith and CTA
  design details (see Le Bohec et al., these proceedings).}

\end{figure}


\section{Science objectives}
Intensity Interferometry at a future IACT is likely to operate in the visual
wavelength region. Science objectives are therefore similar but not the same as
future 1\,km baseline Michelson interferometers such as \emph{Ohana} \cite{2004SPIE.5491..391P} which operate in
the $K$ band. Limiting $\rm m_{v}$ of a CTA concept is illustrated in the right panel
of Fig.\,1. Targets are limited to a $\rm
m_{v}\approx8.5^{m}$ for a S/N = 5, and a 5 hours integration in case of
50\% visibility (see Le Bohec et al., these proceedings). These specifications
allow important interferometric studies regarding binary stars, stellar radii and
pulsating stars with unprecedented resolution on scales of 10s of $\mu$-arcseconds. 
Below we highlight three of the potential science topics.

\paragraph{\bf{Star formation}}
Key questions relating to the physics of mass accretion and pre-main sequence (PMS)
evolution can be addressed by means of very high resolution imaging as provided
by the next generation IACTs and I.I. They involve the 
absolute calibration of PMS tracks, the mass accretion process,
continuum emission variability, and stellar magnetic activity. The technique will be 
able to resolve features on the stellar surface. 
Hot spots deliver direct information regarding the accretion
of material onto the stellar surface. Cool spots on the other hand may cover 50\%
of the stellar surface, and are the product of the slowly decaying rapid
rotation of young stars. Imaging these cool spots will constrain ideas regarding the interplay of
rotation, convection and chromospheric activity and may also
provide a direct practical application in the explanation for the anomalous
photometry observed in young stars \cite{2003AJ....126..833S}.
In practise about 50 young stars with $\rm m_{v}<8^{\rm m}$ are within reach of future
IACTs. In the last decade several young coeval stellar groups have been
discovered within $\sim$50\,pc of the sun. Famous
examples are the TW\,Hydra and $\beta$ Pic comoving groups. The majority of the spectral types
within reach lie between A and G-type. Their ages lie within the range 8
and 50 Myr (for an overview see \cite{2004ARA&A..42..685Z}), 
ensuring that a substantial fraction of the low-mass members are still
in their PMS contraction phase. Measurements of their angular size can be used in
the calibration of evolutionary tracks, fundamental in deriving the properties
of star forming regions and young stellar clusters.  The proximity
of the comoving groups ensures that their members are bright and also makes
these groups relatively sparse, making them unconfused 
targets, even given the large PSF of $\sim3'$. This sparseness is also the
reason for incomplete group memberships, making it likely that the number of
young stars known to lie close to the sun will increase in the years to come.



\paragraph{\bf{Distance scale}}
Measuring the diameters of Cepheids is a basic method with far reaching
implications. A radius estimate of a Cepheid can be obtained using the
Baade-Wesselink method. The Baade-Wesselink method relies on measurement of the
ratio of the star size at times $t_{1}$ and $t_{2}$, based on the luminosity and
colour. Combining this with the measurement of the radial velocity, delivering the
difference in the radius between $t_{1}$ and $t_{2}$, one can derive the radius of the
Cepheid. Combining I.I. angular size measurement with the radius estimate using
the Baade-Wesselink method one obtains the distance to the Cepheid (see
\cite{1994ApJ...432..367S}). This allows the calibration of the all important
Cepheid period-luminosity relation using local Cepheids. A count of Cepheids
observed with Hipparcos \cite{1999A&AS..139..245G} shows that at least 60
Cepheids with $\rm m_{v} <8^{m}$ are within reach of future I.I.-IACTs.

\paragraph{\bf {Rapidly rotating stars}}
As a group classical Be stars are particularly well-known for their proximity to
break-up rotational velocities as deduced from photospheric absorption lines.
In addition these stars show Balmer line emission firmly established as due to
gaseous circumstellar disks, that appear and disappear on timescales of months
to years. These two properties are somehow related, but many open questions
regarding the detailed physical processes at play still exist.
The Be-phenomenon is an important simply due to the number of stars and
the range of stellar physics involved (the fraction of Be stars to normal B-type peaks
at nearly 50\% for B0 stars, \cite{1997A&A...318..443Z}). Absorption lines will
however never provide the final answer to their actual rotational velocity due
to strong gravity darkening at the equator and brightening at the poles.
Direct measurement of the shape of the rotating star is not hampered by
gravity darkening, and provides a direct indication of the rotational speed (see
e.g. $\alpha$ Eri with the VLTI, \cite{2003A&A...407L..47D}).
The Be star disk formation and dissolution activity is little
understood. Photometric observations of Be star disks seem to indicate that they
may actually evolve into ring structures before disappearing into the
interstellar medium (e.g. \cite{2006A&A...456.1027D}). Bremsstrahlung in the disk
can constitute $\sim30\%$ of the total light in $V$-band
\cite{1997A&A...318..443Z}.  
There are about 300 Be stars brighter than $m_{v}=8^m$, roughly
corresponding to a distance limit of 700\,pc \cite{Cat},
signifying that Be star phenomena can be probed in depth with IACT based I.I.

\section{Conclusion}

Intensity Interferometry using the telescopes of the future IACT arrays CTA
and/or AGIS has the potential to make a major impact in several areas of stellar astrophysics
and beyond. Whilst much research and development in this area is still required,
such a major instrument could be realised at rather modest cost and as 
early as 2015.

\bibliographystyle{aipproc}   



\IfFileExists{\jobname.bbl}{}
 {\typeout{}
  \typeout{******************************************}
  \typeout{** Please run "bibtex \jobname" to optain}
  \typeout{** the bibliography and then re-run LaTeX}
  \typeout{** twice to fix the references!}
  \typeout{******************************************}
  \typeout{}
 }

\end{document}